\documentclass[12pt,preprint]{aastex}

\newcommand{\dif}[2]{{\partial #1 \over \partial #2}}
\newcommand{\ve}[1]{\mbox{\boldmath${#1}$}}

\shorttitle{Biermann Mechanism in Primordial Supernova Remnant}
\shortauthors{hanayama et al.}

\begin{document}

\title{Biermann Mechanism in Primordial Supernova Remnant and Seed Magnetic Fields}

\author{Hidekazu Hanayama\altaffilmark{1,2}, Keitaro Takahashi\altaffilmark{3},
Kei Kotake\altaffilmark{4}, Masamune Oguri\altaffilmark{5,6}, \\
Kiyotomo Ichiki\altaffilmark{1,2}, and Hiroshi Ohno\altaffilmark{2,6}}

\altaffiltext{1}{Department of Astronomy, School of Science,
the University of Tokyo, Hongo 7-3-1, Bunkyo, Tokyo, 113-0033, Japan}
\altaffiltext{2}{National Astronomical Observatory of Japan, Mitaka,
Tokyo 181-8588, Japan} 
\altaffiltext{3}{Department of Physics, Princeton University, Princeton,
NJ 08544, U.S.A.}
\altaffiltext{4}{Science \& Engineering, Waseda University, 3-4-1 Okubo,
Shinjuku, Tokyo, 169-8555, Japan}
\altaffiltext{5}{Department of Astrophysical Sciences, Princeton
University, Peyton Hall, Ivy Lane, Princeton, NJ 08544} 
\altaffiltext{6}{Department of Physics, School of Science, University of Tokyo,
7-3-1 Hongo, Bunkyo, Tokyo 113-0033, Japan}

\begin{abstract}
 We study generation of magnetic fields by the Biermann mechanism in
 the supernova explosions of first stars. The Biermann
 mechanism produces magnetic fields in the shocked region between the
 bubble and interstellar medium (ISM), even if magnetic fields are
 absent initially. We perform a series of two-dimensional
 magnetohydrodynamic simulations with the Biermann term and estimate the
 amplitude and total energy of the produced magnetic fields. We find
 that magnetic fields with amplitude $10^{-14}-10^{-17}$ G are generated
 inside the bubble, though the amount of magnetic fields generated
 depend on specific values of initial conditions. This corresponds to
 magnetic fields of $10^{28}-10^{31}$ erg per each supernova remnant,
 which is strong enough to be the seed  magnetic field for galactic
 and/or interstellar dynamo.   
\end{abstract}


\section{Introduction}

Magnetic fields are ubiquitous in the universe. In fact, observations of
rotation measure and synchrotron radiation have revealed that magnetic
fields exist in astronomical objects with various scales: galaxies,
clusters of galaxies, extra-cluster fields, etc (for a review, see
e.g., \citealt{wid02}). The observed galactic magnetic fields have
both coherent and fluctuating components whose strengths are comparable
to each other \citep{fos02,han04}. Conventionally these magnetic fields
are considered to be amplified and maintained by dynamo mechanism. The
coherent component in a galaxy is expected to be amplified by galactic
dynamo, while the fluctuating component may be amplified by interstellar
dynamo driven by turbulent motion of the interstellar medium (ISM)
\citep{bal04}. However, the dynamo mechanism itself cannot explain the
origin of the magnetic fields: The seed magnetic fields are needed for 
the dynamo mechanism to work.

Thus far various mechanisms have been suggested as a possible source of
the seed field. They can be classified into two types, astrophysical and
cosmological origins. Here we concentrate on the former (for the latter
scenario,  see e.g., \citealt{les95,dav00,bam04,ber04,yam04,tak05}). 
Basically, the astrophysical magnetogenesis invokes the Biermann
mechanism \citep{bie50} which is induced by the electric currents
produced by the spatial gradient of the electron pressure not parallel to
that of the density. This is a pure plasma effect so that there is no
need to assume unknown physics as is often done in cosmological
models. Because the Biermann mechanism requires the non-parallel spatial
gradient of the pressure and density, some non-adiabatic process is
necessary to produce deviation of the equation of state from the
polytropic one. The strong magnetic fields in high-redshift galaxies
\citep{ath98} imply that the significant amount of the seed magnetic
field should be generated at early stage, e.g., epoch of cosmological
reionization or protogalaxy formation. For instance, \citet{gne00}
studied the generation of magnetic fields in the ionizing front, and
found that magnetic fields as high as $\approx 10^{-18}$ G in virialized
objects can be generated. \citet{kul97} showed that magnetic filed of
$\approx 10^{-21}$ G can also be generated at shocks of large-scale
structure formation. 

In this paper, we investigate magnetogenesis at smaller
scales. Specifically, we study the amplitude of the magnetic field
produced by the Biermann mechanism when the shock waves of the
supernova explosions of the first stars are spreading throughout the ISM.
The primordial supernova explosions are expected to take place
effectively, since the initial mass function (IMF) of population III
stars should be substantially top-heavy \citep[e.g.,][]{abe02}.
We perform a series of the two dimensional magnetohydrodynamic (MHD)
numerical simulations in which the Biermann term is included.  We also
discuss whether they can be the origin of the cosmic magnetic
fields. Consequently, we find that the spatially-averaged amplitude
of the produced magnetic field in virialized objects reaches $\sim 10^{-16}$ G,
which is much greater than those expected from cosmic reionization
and large-scale structure formation. Thus the supernova explosions of the
first stars can be effective sources for the seed magnetic fields.
Although the situation considered here is somewhat similar to
that of \citet{mir98}, they assumed multiple explosion scenario of
structure formation and considered explosions of objects with mass
$>10^6M_\odot$ at $z\gtrsim 100$, which is clearly unrealistic in the
context of the current standard model of structure formation.

\section{Numerical Simulations}

\subsection{Numerical Method}
The basic equations for our numerical simulations are as follows:
\begin{equation}
\frac{\partial\rho}{\partial t} + \nabla \cdot (\rho \mathbf{v}) = 0
\label{mass}
\end{equation}
\begin{equation}
\frac{\partial\rho \mathbf{v}}{\partial t} + \nabla \cdot 
(\rho \mathbf{v} \mathbf{v}) = - \nabla P - \nabla 
(\frac{{\mathbf{B}}^{2}}{8\pi})
+ \frac{1}{4\pi} (\mathbf{B} \cdot \nabla) \mathbf{B}
\label{mome}
\end{equation}
\begin{equation}
\frac{\partial}{\partial t} (E + \frac{{\mathbf{B}}^{2}}{8\pi})
+ \nabla \cdot \left[ (E + P) \mathbf{v}+ \frac{1}{4\pi} 
\left\{ \mathbf{B} \times (\mathbf{v} \times \mathbf{B}) \right\} 
\right]
= - \Lambda + \Gamma
\label{ener}
\end{equation}
\begin{equation}
\frac{\partial \mathbf{B}}{\partial t}
= \nabla \times (\mathbf{v} \times \mathbf{B})
  + \alpha \frac{\nabla \rho \times \nabla P}{\rho^{2}}
\label{ind}
\end{equation}
where $\rho$, $\mathbf{v}$, $e$, $P$, and $\mathbf{B}$ are density,
velocity, total energy, pressure, and magnetic field, respectively. Here
the total energy of the gas is defined as, 
\begin{equation} 
E = \frac{1}{2}\rho v^2 + \frac{P}{\gamma - 1},
\end{equation}
where we fix the value of the adiabatic index $\gamma$ to $5/3$,
which is valid in case of the non-relativistic gas. The last term of the
right hand side of equation (\ref{ind}) is the Biermann term:  $\alpha$ 
in equation (\ref{ind}) is the so-called Biermann coupling constant
defined by $\alpha = m_{p} c/e (1 + \chi) \sim 10^{-4}~{\rm G \cdot s}$,
where $m_{p}$, $e$, and $\chi$ being the proton mass, electric charge,
and ionization fraction, respectively. Although the gas temperature just
before the star formation is rather cool $T \sim 200 {\rm K}$
\citep{abe02,bro03,omu03}, UV radiation from the first stars ionizes the
surrounding ISM \citep{fre03,mor04}. Thus we set $\chi=5/6$ assuming 
$n_{He}/n_{H} \sim 0.1$. The radiative cooling is represented by
$\Lambda$ in equation (\ref{ener}) and we use a cooling function derived
by \citet{ray76}. When parcels cool below $10^{4}$K, an artificial
heating rate proportional to the density, $\Gamma$ in equation (\ref{ener}),
is used. The constant heating coefficient is set so that heating
balances cooling at the ambient density and temperature. Although the
cooling function and heating rate in the primordial gas are not clear so
far, they are not important for the adiabatic expansion phase which we
concentrate on. 

We solve the above equations by the two-dimensional MHD code in the
cylindrical coordinates $(r, z, \phi)$ assuming axial symmetry around
the symmetry axis ($z$). The code is based on the modified Lax-Wendroff
scheme with an artificial viscosity of von Neumann and Richtmyer to
capture shocks. The numerical scheme was tested by comparing known
solutions which have been obtained either analytically and numerically.  
Specifically, we checked the code by comparing with [1] the adiabatic SNR
with the Sedov solution \citep{sed59}, [2] the spherically symmetric SNR
\citep{sla92} and [3] MHD shock tube problems \citep{bri88}.  
The analytical solution of [1] is reproduced within 5\% relative error,
and the oscillations behind the shock front are well suppressed. 

The artificial viscosity is added to smooth the jump at the shock front 
in equations (\ref{mass}) - (\ref{ind}). 
For all the quantities $U\equiv$ ($\rho$, $\rho\ve{v}$, $E +
{\mathbf{B}}^{2}/8\pi$, \ve{B}),  
we added an extra term to express artificial diffusion and solved equations like
\begin{equation}
\dif{U}{t}=\nabla(c \nabla U),
\end{equation}  
where we take 
\begin{equation}
 c=A_v\left[\left(\dif{v_z}{z}\right)^2+\left(\dif{v_z}{r}\right)^2
+\left(\dif{v_z}{z}\right)^2+\left(\dif{v_r}{r}\right)^2\right]^{1/2}\Delta^2, 
\end{equation}
with $\Delta=\Delta z=\Delta r$ and $A_v=4$.

In all the computations, grid spacings are chosen $\Delta r=\Delta z=0.1$ pc.
For example, numerical domain covers a region of 130 pc $\times$ 130 pc 
with 1300 ($r$) $\times$ 1300 ($z$) mesh points in our fiducial model 
(see  $\S$\ref{sec:2.2}). Even if the resolution is doubled, the total energy 
of magnetic field is not changed 4 times at $t = 10^5$ yrs.
We begin the simulation by adding thermal
energy of $E_0=10^{53}$ or $ 10^{52}{\rm erg}$ within the sphere of 2pc
in radius.   

\subsection{Results of Numerical Simulations}
\label{sec:2.2}
As the bubble expands, the ejected gas interacts with ISM and shock wave
is formed. In the shocked region the gas is heated non-adiabatically,
which is a necessary condition for the Biermann mechanism to work. Here
the structure of the interstellar environment is important because it
affects the density and pressure profiles of the shocked region which is
directly related to the Biermann term. We assume inhomogeneous  ISM with
average density $n_{\rm ISM} = 0.2 {\rm cm}^{-3}$. This is roughly
consistent with the situation discussed in \citet{bro03}. The scale
length of the density and the amplitude of the inhomogeneity are poorly
understood now and we assume inhomogeneity with the scale length
$\lambda = 1 {\rm pc}$ and density variation $0.2 \times 2^{\pm 1} {\rm
cm}^{-3}$, which are similar values to those in our galaxy. 
Within the variation, the amplitude of density is given at random 
and the distribution is smoothed numerically to perturb. 
This is our fiducial model for the ISM. As we will show in the next section, 
the amplitude of  the produced magnetic field is sensitive to scale length
$\lambda$, while the average density and the amplitude of the density
variation is rather unimportant. Thus we consider several different
models in addition to the fiducial model: Specifically, we vary the mean
density ($1 \times 2^{\pm 1} {\rm cm}^{-3}$ and $10 \times 2^{\pm 1} {\rm
cm}^{-3}$) and the scale length ($3 {\rm pc}$ and $10 {\rm pc}$). 

As for the explosion energy of the supernova, we adopt $E_{\rm SN} =
10^{53} {\rm erg}$ for the fiducial mode. This explosion energy
corresponds to stars with mass $250 M_{\odot}$ which explodes as a
pair-instability supernova \citep{fry01}. Also we consider a model with
$E_{\rm SN} = 10^{52} {\rm erg}$ as a variation. 

\begin{figure}
\begin{center}
\includegraphics[width=7.5cm,clip]{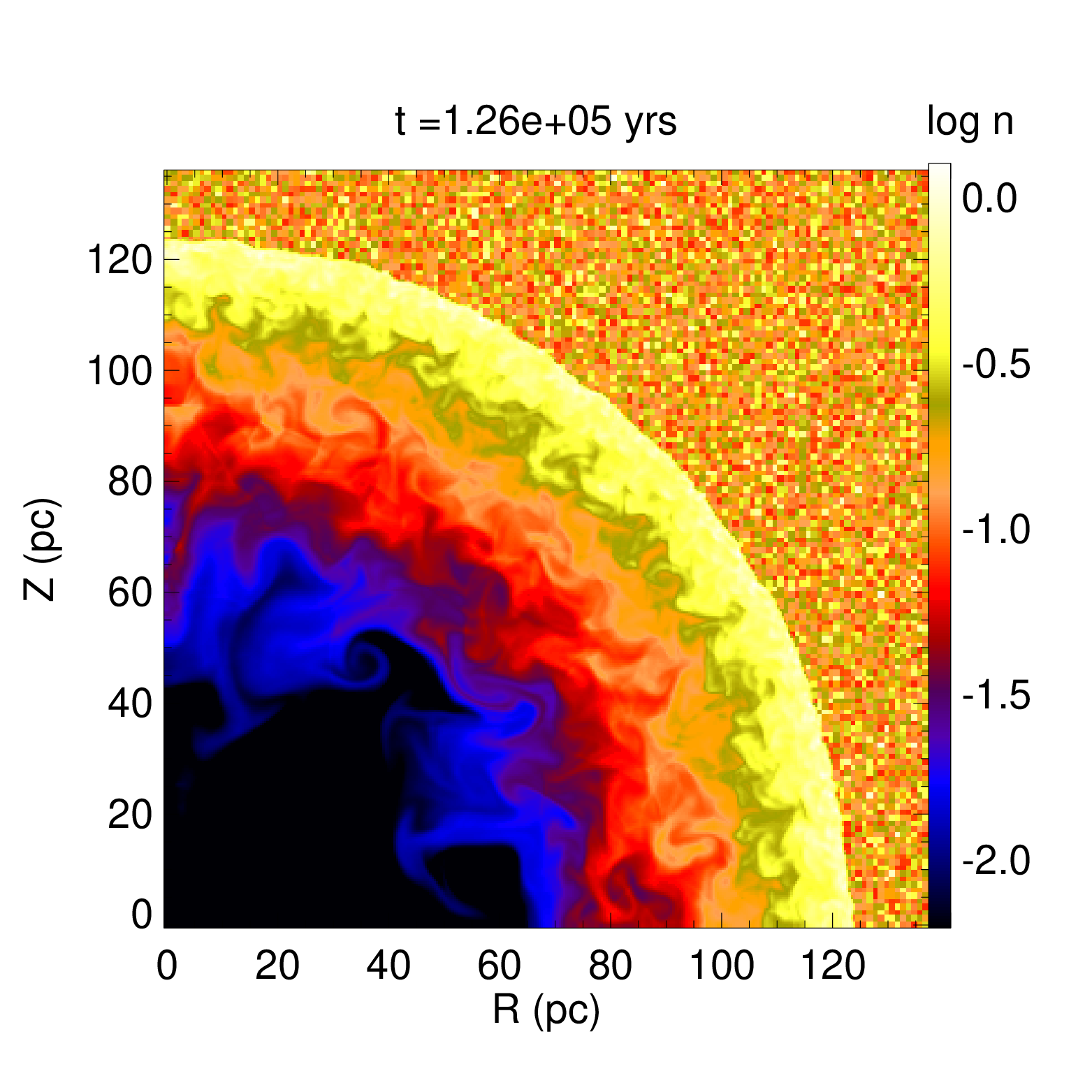}
\includegraphics[width=7.5cm,clip]{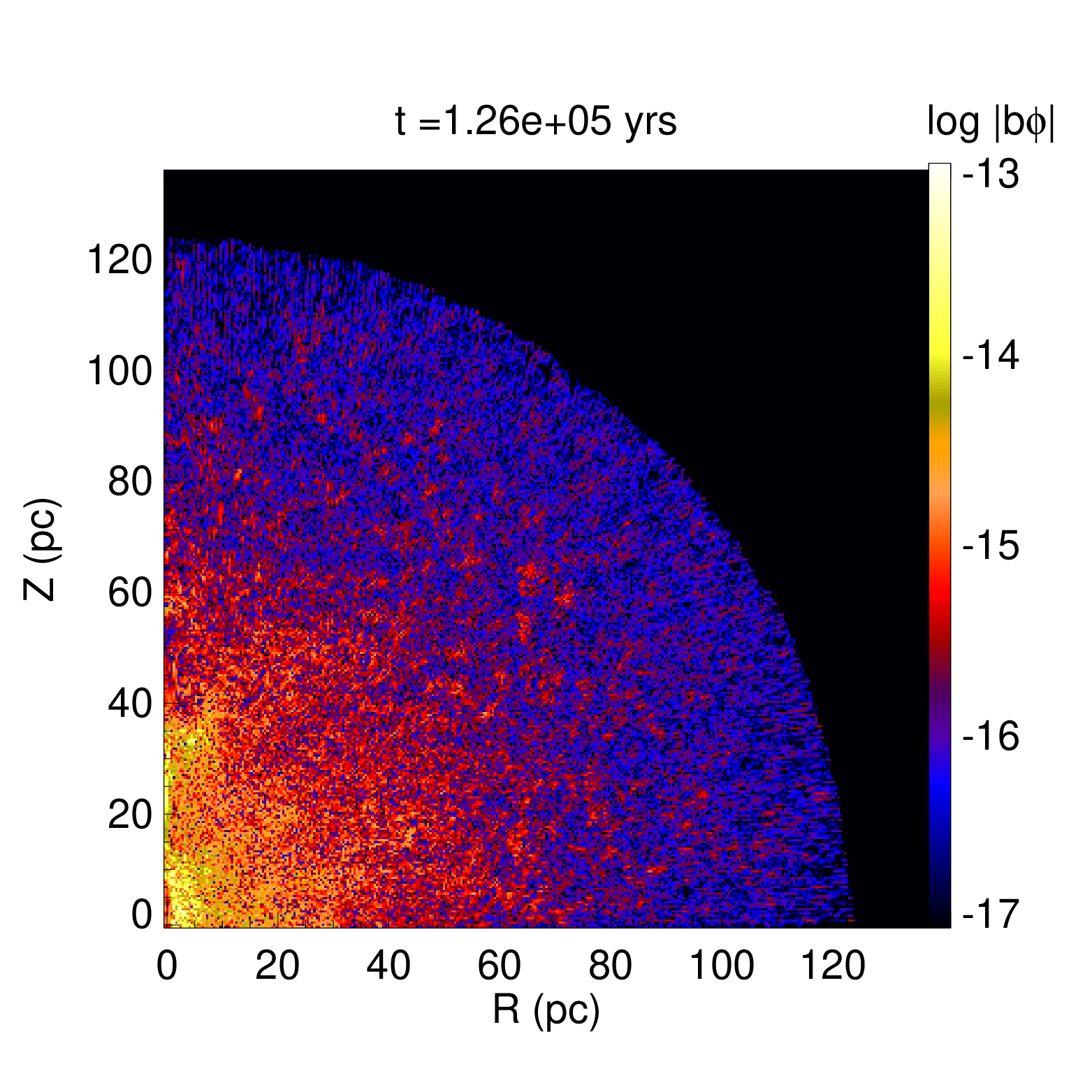}
\end{center}
\caption{Contours of the gas density (left) and the magnetic field (right) at the end of
the adiabatic expansion phase ($t = 1.26 \times 10^{5} {\rm yr}$) for the fiducial model.
The amplitude of the magnetic field is about $10^{-14} {\rm G}$ at the central region
and about $10^{-17} {\rm G}$ just behind the shock.
\label{fig:contour}}
\end{figure}

Figure \ref{fig:contour} shows the contours of the gas density and
amplitude of the produced magnetic field at the end of the adiabatic
expansion phase $t = 1.26 \times 10^{5} \rm{yrs}$ for the fiducial
model. The radius of the bubble is about 125 pc and turbulent motion is
induced in the shocked region due to the inhomogeneity of the ISM. The
amplitude of the magnetic field is about $10^{-14} {\rm G}$ for the
central region and about $10^{-17} {\rm G}$ just behind the shock. 
The total magnetic energy inside the bubble is about $10^{30} {\rm erg}$. 

We have also checked the case of homogeneous medium to test robustness of
our computation. We find that SNR generates the magnetic fields with
amplitude $\sim10^{-19}$ G behind the shock front in average, and it is
$\sim10^{-3}$ times smaller than that in inhomogeneous medium. Therefore,
in case of homogeneous medium, we estimate a numerical error of the
amplitude of magnetic field as $\sim10^{-19}$ G.  It should be noted
that the produced magnetic field has only toroidal component because
axial symmetry was assumed.  

\begin{figure}
\begin{center}
\includegraphics[width=15cm,clip]{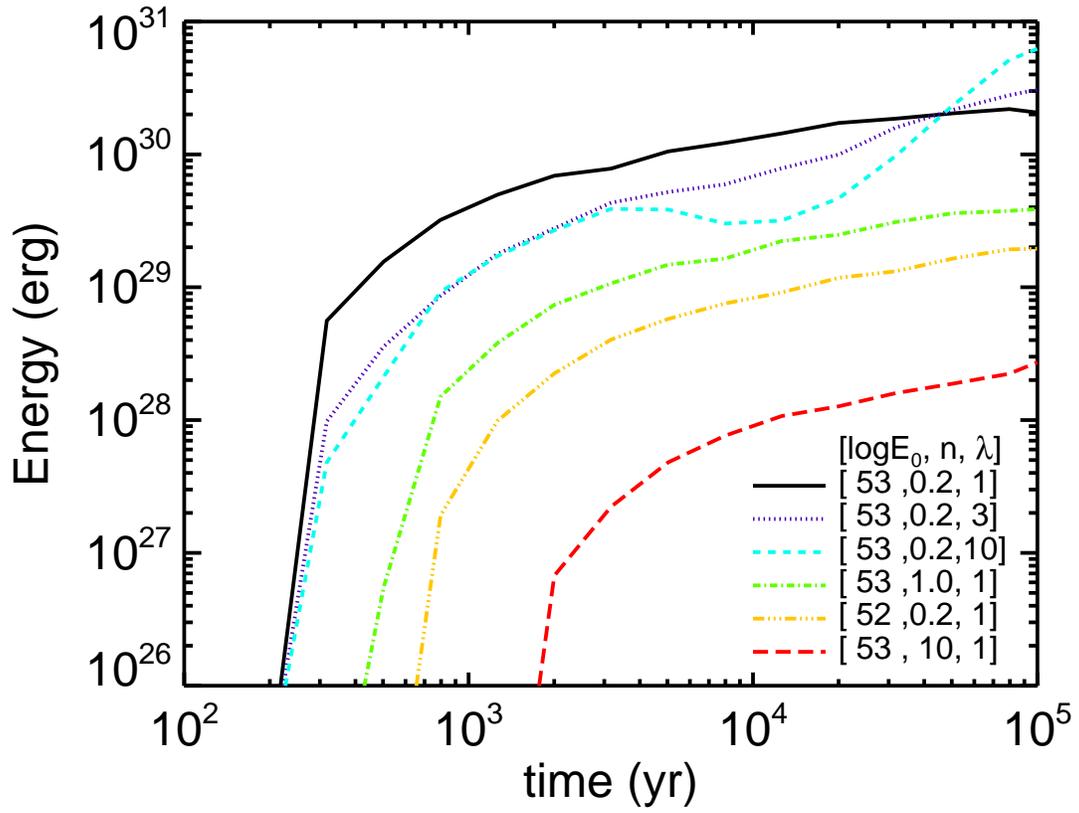}
\end{center} 
\caption{Time evolution of the total magnetic energy produced by the
 Biermann mechanism for various models.
\label{fig:evolution}}
\end{figure}

In Figure \ref{fig:evolution}, we show the time evolution of the total
magnetic energy for various models. The behaviors are qualitatively
similar for all the models. The robust knees around $10^3$ years in
Figure \ref{fig:evolution} come from a formation of adiabatic shock
front: It corresponds to when SNR shifts from free expansion phase to
Sedov phase. The total magnetic energy at the end of the adiabatic
expansion phase  is larger for models with smaller scale length of the
ISM density.  This tendency will be confirmed by an order-of-magnitude
estimate in the next section. For models with large average ISM density
or small explosion energy,  the total magnetic energy is smaller because
the bubble is smaller than  the other models. Although there are many
uncertainties in initial conditions,  the generation of magnetic field
with the total energy  $10^{28}-10^{31} {\rm erg}$ appears to be robust.

\section{Analytic Estimates of Magnetic Fields in Primordial Supernova
 Remnants and Implications for the Seed Magnetic Field}

To understand the result of numerical simulations, in this section we
perform an order-of-magnitude estimation of the strength of the magnetic
field produced by the Biermann mechanism. 

The amplitude of the magnetic field produced by the Biermann mechanism can
be estimated from the Biermann term in equation (\ref{ind})
\begin{equation}
B_{\rm Biermann} \sim \alpha \frac{\nabla \rho \times \nabla P}{\rho^{2}} \Delta t,
\end{equation}
where $\Delta t \sim 10^{3}~{\rm yr}$ is a characteristic timescale the
Biermann mechanism works. Taking the characteristic pressure to be the
ram pressure of the gas, $P \sim P_{\rm ram} = \rho~v_{\rm bubble}^{2}$,
and the characteristic velocity  of the bubble $v_{\rm bubble} \sim
10^{-3}~{\rm pc\,yr^{-1}}$, we obtain
\begin{eqnarray}
B_{\rm Biermann}
&\sim& \alpha \frac{v_{\rm bubble}^{2}}{\lambda L} \Delta t \nonumber \\
&\sim& 3 \times 10^{-15}
       \left(\frac{v_{\rm bubble}}{10^{-3} {\rm pc\,yr^{-1}}}\right)^{2}
       \left(\frac{\lambda}{1{\rm pc}}\right)^{-1}
       \left(\frac{L}{1{\rm pc}}\right)^{-1}
       \left(\frac{\Delta t}{10^{3}{\rm yr}}\right)~{\rm G},
\end{eqnarray}
where $L$ is a scale length of the pressure component perpendicular to
the density gradient. Then the magnetic energy produced by the Biermann
mechanism per each primordial supernova remnant can be estimated as
\begin{eqnarray}
E_{B} &\sim& 
\frac{4 \pi R_{\rm bubble}^{3}}{3} \frac{B_{\rm Biermann}^{2}}{8 \pi} 
\nonumber \\
&\sim& 5 \times 10^{31}
       \left(\frac{R_{\rm bubble}}{100{\rm pc}}\right)^{3}
       \left(\frac{v_{\rm bubble}}{10^{-3} {\rm pc\,yr^{-1}}}\right)^{4}
       \left(\frac{\lambda}{1{\rm pc}}\right)^{-2}
       \left(\frac{L}{1{\rm pc}}\right)^{-2}
       \left(\frac{\Delta t}{10^{3}{\rm yr}}\right)^2 {\rm erg},\label{eb}
\label{eq:E-estimate}
\end{eqnarray}
which is consistent with the value obtained from our numerical simulations.

The dependence of the total magnetic energy on several parameters can 
also be understood from equation (\ref{eb}). It is found that $E_B\propto
\lambda^{-2}$ directly from equation (\ref{eb}). To examine the
dependence on the other parameters, we simply assume the Sedov-Taylor
solution: 
\begin{equation}
R_{\rm bubble} \propto t^{2/5}\left(E_{\rm SN}/n_{\rm ISM}\right)^{1/5}, 
\;\;\;\; 
v_{\rm bubble} \propto t^{-3/5}\left(E_{\rm SN}/n_{\rm ISM}\right)^{1/5}.
\end{equation}
Putting these into equation (\ref{eb}) yields 
$E_B\propto(E_{\rm SN}/n_{\rm ISM})^{7/5}$. Our numerical results shown
in Figure \ref{fig:evolution} are quite consistent with these simple
estimations. 

Now we estimate the spatially-averaged energy density of the magnetic
fields produced by the first stars and consider if they can be a source of
the seed fields. As for the primordial star formation rate, we
extrapolate the one by \citet{pel04} and \citet{ric04};  
$\dot{\rho_{\star}} \sim 10^{-2}~M_{\odot}~{\rm yr}^{-1} {\rm Mpc}^{-3}$.
Denoting the magnetic energy produced by the Biermann mechanism as
$\epsilon_{SN} \sim 10^{30}{\rm erg}$, the magnetic energy density
produced during the formation period of the first-star ($\tau \sim
1~{\rm Gyr}$) can be obtained as 
\begin{eqnarray}
e_{\rm B} & \sim & f_{\gamma\gamma}\dot{\rho_{\star}}~
\Bigl(\frac{\epsilon_{SN}}{M_{\rm SN}}\Bigr)\tau \nonumber \\
&\sim& 10^{-40}
       \left(\frac{f_{\gamma\gamma}}{0.06}\right)
       \left(\frac{\dot{\rho_{\star}}}{10^{-2}~M_{\odot}~{\rm yr}^{-1}}\right)
       \left(\frac{M_{\rm SN}}{250 M_{\odot}} \right)^{-1}
       \left(\frac{\epsilon_{SN}}{10^{30}{\rm erg}}\right)
       \left(\frac{\tau}{1~{\rm Gyr}}\right) {\rm erg\,cm^{-3}},
\label{eq:10}
\end{eqnarray}
where $M_{\rm SN}$ is the typical mass scale of first
stars that end up in pair instability supernovae, and $f_{\gamma\gamma}$
is the mass fraction of such first stars: We adopt
$f_{\gamma\gamma}=0.06$ that was  derived under the assumption that very
massive black holes produced from first stars end up in supermassive
black holes in galactic centers \citep{sch02}. We note that the value is the
comoving density averaged in the universe: We can convert the value to
physical density in virialized objects (i.e., protogalaxies) as
\begin{equation}
 e_{\rm B,gal}\sim e_{\rm B}(1+z)^4\Delta \sim 10^{-34}
\left(\frac{e_{\rm B}}{10^{-40}{\rm erg\,cm^{-3}}}\right)
\left(\frac{1+z}{10}\right)^4
\left(\frac{\Delta}{200}\right) {\rm erg\,cm^{-3}},
\label{eq:11}
\end{equation}
where $\Delta$ is the density contrast. This corresponds to the mean
magnetic field of $B\sim 10^{-16}$ G in protogalaxies, which is much
stronger than expected in ionizing fronts, $B \sim 10^{-18}$ G
\citep{gne00}. 

\section{Summary and Discussions}
\label{sec:4}

We have studied the generation of magnetic fields in primordial
supernova remnants. We have performed the two dimensional MHD simulations
with the Biermann term which can produce magnetic field through
the non-adiabatic interaction between the bubble and ISM, even if there
is no magnetic field at first. We have found that the ISM around the
primordial supernovae is an effective site to produce magnetic fields. 
The the total energy of the magnetic fields is $10^{28}-10^{31} {\rm
erg}$, depending on parameters adopted. Based on the results, we have
estimated spatially-averaged energy density of the magnetic fields
produced by the first stars during the formation period of the first
stars. The averaged energy density is about $10^{-40} {\rm erg
~cm}^{-3}$, which corresponds to $B\sim 10^{-16}$ G in protogalaxies at
$z\sim 10$.  This is much greater than expected from cosmic reionization
and large-scale structure formation. Thus primordial supernova remnants would
be a promising source for the seed fields for galactic and/or
interstellar dynamo.  

We note that our results are based on two-dimensional MHD
simulation. This mechanism may behave very differently in more
realistic three dimensional case, primarily because the vorticity
cascade may be different. We are preparing for the three-dimensional
MHD simulations to study this possible difference, which will be
presented elsewhere.

Although the coherence length of the seed field computed here is much
smaller than the galactic scale, it can be amplified by the galactic
dynamo to produce the coherent component if the coherence length is
about 100 pc \citep{poe93,bec94,fer00}, which is a typical size of
supernova remnants. It might be also amplified by interstellar dynamo to
produce the fluctuating component \citep{bal04}. While it is beyond our
scope to discuss the relation between the produced magnetic field and
galactic/interstellar dynamo, we plan to investigate the evolutions of
the seed magnetic fields computed here as a result of the dynamo
processes in the large scale of galaxies and cluster of galaxies. This
will be presented in the forthcoming paper (Hanayama et al. in
preparation).  

As observational signature, a proposal to detect seed magnetic fields
was given by e.g., \citet{pla95}. If there exists intergalactic magnetic
fields produced by the primordial SNR,  an arrival time of high energy
gamma-ray photons from extragalactic would be delayed by the action of
intergalactic magnetic fields on electron cascades \citep{and04}.  The
magnetic field as weak as $\sim10^{-24}$ G would be detectable if the
delay time comes into reasonable range, a few days.  Therefore the
mechanism of magnetic field generation proposed in this paper might be
tested in the future high-energy gamma-ray experiments such as GLAST. 
 
\acknowledgments
We would like to thank Kohji Tomisaka for invaluable comments and advice 
on writing the manuscript. We are grateful to Kanako Sugimoto and Hiromi 
Mizusawa for useful comments. Numerical computations were carried out on
the VPP5000 supercomputer  at the Astronomical Data Analysis Center of
the National Astronomical Observatory, Japan. We also thank Ryoji
Matsumoto and Takaaki Yokoyama for a contribution to the calculation
code, CANS(Coordinated Astronomical Numerical Software).  K. T., K. K.,
M. O. and K. I. are supported by Grant-in-Aid for JSPS Fellows. 

\end{document}